\RequirePackage{fix-cm}
\documentclass[smallextended]{svjour3}       
\smartqed  
\usepackage{graphicx}
\begin{document}

\title{Multi-wave, breather wave and lump solutions of the
Boiti-Leon-Manna-Pempinelli equation with variable
coefficients\thanks{This research was supported by the National Natural Science Foundation of China(No. 61762051), the Jiangxi province science and technology plan projects (No. 20171BBG70108), the University level project of Jiangxi University of traditional Chinese medicine(Big dataanalysis and integration platform of traditional Chinese medicine intelligent manufacturing system)}}


\author{ Jian-Guo Liu$^{*}$,  Wang-Ping Xiong$^{*}$
}


\institute{Jian-Guo Liu(*Corresponding author)\at
              College of Computer, Jiangxi University of
Traditional Chinese Medicine, Jiangxi 330004, China\\
              Tel.: +8613970042436\\
              \email{20101059@jxutcm.edu.cn}           
\and Wang-Ping Xiong(*Corresponding author)\at
College of Computer, Jiangxi University of
Traditional Chinese Medicine, Jiangxi 330004, China\\
\email{20030730@jxutcm.edu.cn}  }

\date{Received: date / Accepted: date}

\maketitle

\begin{abstract}
In this paper, a variable-coefficient Boiti-Leon-Manna-Pempinelli
equation is to be investigated. We obtain abundant multi-wave,
breather wave and lump solutions  by using the three waves method,
 the homoclinic breather
approach and the Hirota's bilinear method. All solutions have been
verified to be correct with the help of Mathematica software. We
study the propagation characteristics of these solutions by some
three-dimensional images. The obtained results are useful for
understanding fluid propagating and  incompressible fluid.

\end{abstract}

\section{Introduction}
\label{intro}\quad With the development of symbolic computation,
nonlinear evolution equations (NLEE) and their solutions play an
important role in almost all branches of physics, such as
hydrodynamics, plasma physics, nonlinear quadratic media, optical fiber [1-9]. Then various
methods were proposed for finding exact solutions of NLEE by the
researchers, such as wave function ansatz method [10-12], three-wave approach [13], Hirota bilinear method
[14], Darboux Transformations [15], linear superposition principle
[16], Riemann-Hilbert approach [17], homoclinic breather approach
[18], homogeneous balance method [19], G'/G-expansion method [20],
Bell polynomial approach [21, 22], and so on. NLEE with variable
coefficients are often able to describe more complex physical
phenomena, so they have attracted great attention, such as Hirota
equation with variable coefficients [23], (2+1)-dimensional
nonlinear Schr\"{o}dinger equation with variable coefficients [24],
Kadomtsev-Petviashvili equation (KP) with variable coefficients
[25], (2+1)-dimensional nonlinear Heisenberg ferromagnetic spin
chain equation [26], variable-coefficient Korteweg-de Vries equation
[27, 28], fifth-order variable-coefficient Sawada-Kotera equation
[29], generalized KP equation with variable coefficients [30, 31],
Bogoyavlensky-Konopelchenko equation [32], and so on. Inspired by
these literatures,  a (2+1)-dimensional variable-coefficient
Boiti-Leon-Manna-Pempinelli equation (vcBLMPe) is studied as follows
[33]
\begin{eqnarray}
a(t) u_{xt}+b(t) u_{yt}+c(t) u_{xy}+k(t) u_{yy}+3
   \left(u_x u_{xy}+u_y
   u_{xx}\right)+u_{xxxy}=0,
\end{eqnarray}where $u=u(x,y,t)$,
$a(t), b(t), c(t)$ and $k(t)$ are differentiable functions. Soliton
solution and new bilinear B\"{a}cklund transformation by using Bell
polynomial technique and bilinear method have been obtained in Ref .
[33]. As far as we know, multi-wave, breather wave and lump
solutions of Eq. (1)
 have not been studied yet.

\quad By using Bell polynomial technique, the bilinear form of
vcBLMPe can be written as{\begin{eqnarray} (D_y
D_x^3+a(t)D_t\,D_x+b(t)D_t\,D_y+c(t)D_y\,D_x+k(t) D_y^2) \xi\cdot
\xi=0
\end{eqnarray}
}with  $u=2\,[ln\xi(x,y,t)]_x$, which is equivalent to
\begin{eqnarray} &&\xi  [a(t) \xi_{xt}+b(t) \xi_{ty}+c(t) \xi_{xy}
+k(t) \xi_{yy}+\xi_{xxxy}]-a(t) \xi_t \xi_x\nonumber\\&&-b(t) \xi_t
\xi_y-c(t)
   \xi_x \xi_y-k(t) \xi_y^2-\xi_{xxx}
   \xi_y+3 \xi_{xy} \xi_{xx}-3 \xi_x \xi_{xxy}=0.
\end{eqnarray}

\quad The structure of this paper is as follows. Section 2 presents
multi-wave solutions by using the three waves method with variable
coefficients; Section 3 obtains the breather wave solutions based on
the homoclinic breather approach; Section 4 studies the lump
solutions by applying the Hirota's bilinear method with variable
coefficients; Section 5 gives the conclusions.

\section{Multi-wave solutions} \label{sec:2}
\quad Multi-wave solutions have been studied for a long time, such as the  $X$ waves,  originally found in acoustics, have been proposed as a new
paradigm in areas ranging from classical to quantum optics [34]. In order to search the multi-wave solutions of Eq. (1), we
directly assume a solution consisting of three different types of
functions (named three waves method [35]) as {\begin{eqnarray}
\xi&=&\theta _2(t) \cos [\varphi _6(t)+\varphi _4 x+\varphi _5
y]+\theta _1(t) \sinh
   [\varphi _3(t)+\varphi _1 x+\varphi _2 y]\nonumber\\&+&\theta _3(t) \cosh [\varphi
   _9(t)+\varphi _7 x+\varphi _8 y],
\end{eqnarray}}where  $\varphi_i(1\leq i \leq 9)$ and $\theta_i(t) (i=1,2,3)$ are
unknown parameters. Substituting Eq. (4) into Eq. (3), we can derive
the following multi-wave solutions of Eq. (1)
\begin{eqnarray}u&=&[2 [-\varphi _4 \chi _1 \sin [\varphi _6(t)+\varphi _4 x-\frac{\varphi _2 \varphi
   _7^2 y}{\varphi _1 \varphi _4}]+\varphi _7 \chi _2 \sinh [\varphi _9(t)+\varphi _7
   (x+\frac{\varphi _2 y}{\varphi _1})]\nonumber\\&+&\varphi _1 \cosh [\varphi
   _3(t)+\varphi _1 x+\varphi _2 y]]]/[\chi _1 \cos [\varphi _6(t)+\varphi _4
   x-\frac{\varphi _2 \varphi _7^2 y}{\varphi _1 \varphi _4}]\nonumber\\&+&\chi _2 \cosh [\varphi
   _9(t)+\varphi _7 (x+\frac{\varphi _2 y}{\varphi _1})]+\sinh [\varphi
   _3(t)+\varphi _1 x+\varphi _2 y]].
\end{eqnarray} All parameters have been interpreted in Appendix A.

\quad   For the sake of  analyzing the physical structure of the
multi-wave solutions more  concretely, two illustrated examples are
given owing to the existence of variable coefficients in Eq. (1).

\quad First, we choose
\begin{eqnarray}a(t)&=&c(t)=\chi_4=1,
\varphi_1=3, \varphi_7=-1, \varphi_2=\varphi_5=-2,\nonumber\\
\chi_3&=&\chi_5=\varphi_4=2, \chi_1=4,
 \epsilon_1=1\nonumber.
\end{eqnarray}Then, the corresponding multi-wave solution can be
read as
\begin{eqnarray}u&=&[-80 \sin \left(-\frac{412 t}{5}+2 x-2 y+1\right)-6 \sqrt{65} \sinh \left(-\frac{424
   t}{5}-x-\frac{2 y}{13}+2\right)\nonumber\\&+&30 \cosh \left(-\frac{472 t}{5}+3 x-2 y+2\right)]/[20 \cos
   \left(-\frac{412 t}{5}+2 x-2 y+1\right)\nonumber\\&+&5 \sinh \left(-\frac{472 t}{5}+3 x-2 y+2\right)+3
   \sqrt{65} \cosh \left(-\frac{424 t}{5}-x-\frac{2 y}{13}+2\right)].
\end{eqnarray}Fig. 1(a), Fig. 1(b) and Fig. 1(c) describe the propagation characteristics of Eq. (6) at  $t = 0$,
$x = 0$ and $y = 0$, respectively.   The $X$ solitary wave fission can be observed in Fig. 1(b) and Fig. 1(c). The interaction between soliton and lump wave is shown in Fig. 1(a).

\includegraphics[scale=0.4,bb=20 270 10 10]{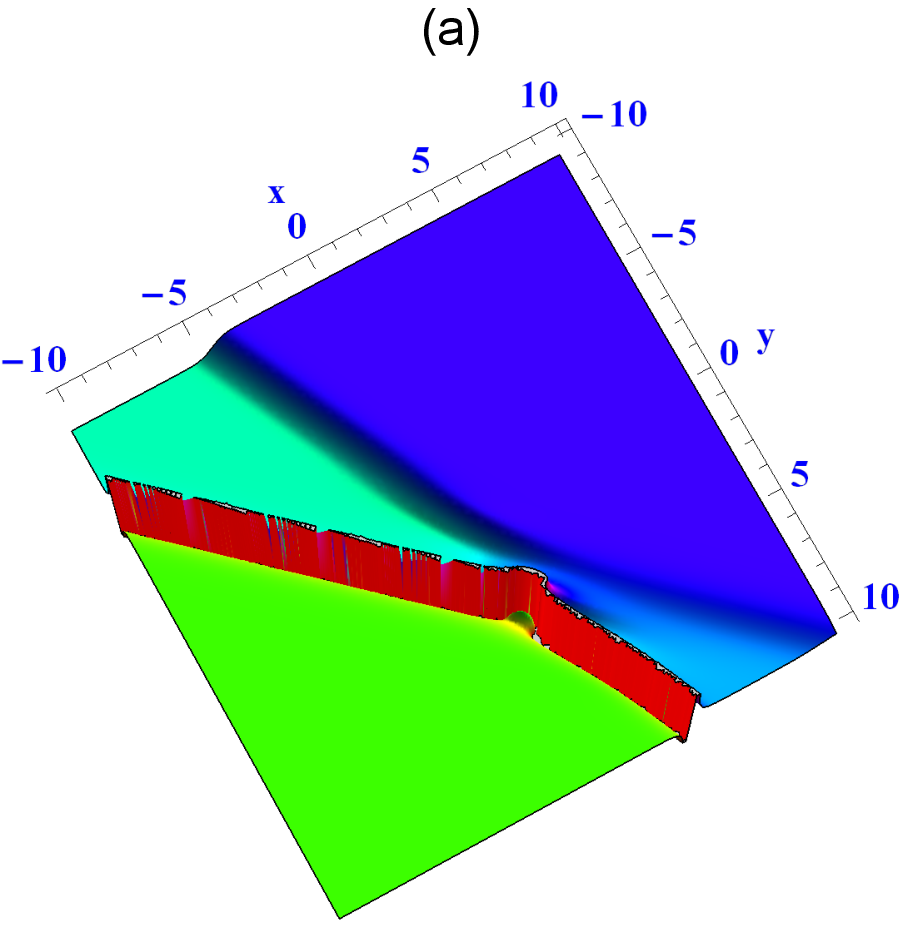}
\includegraphics[scale=0.4,bb=-255 270 10 10]{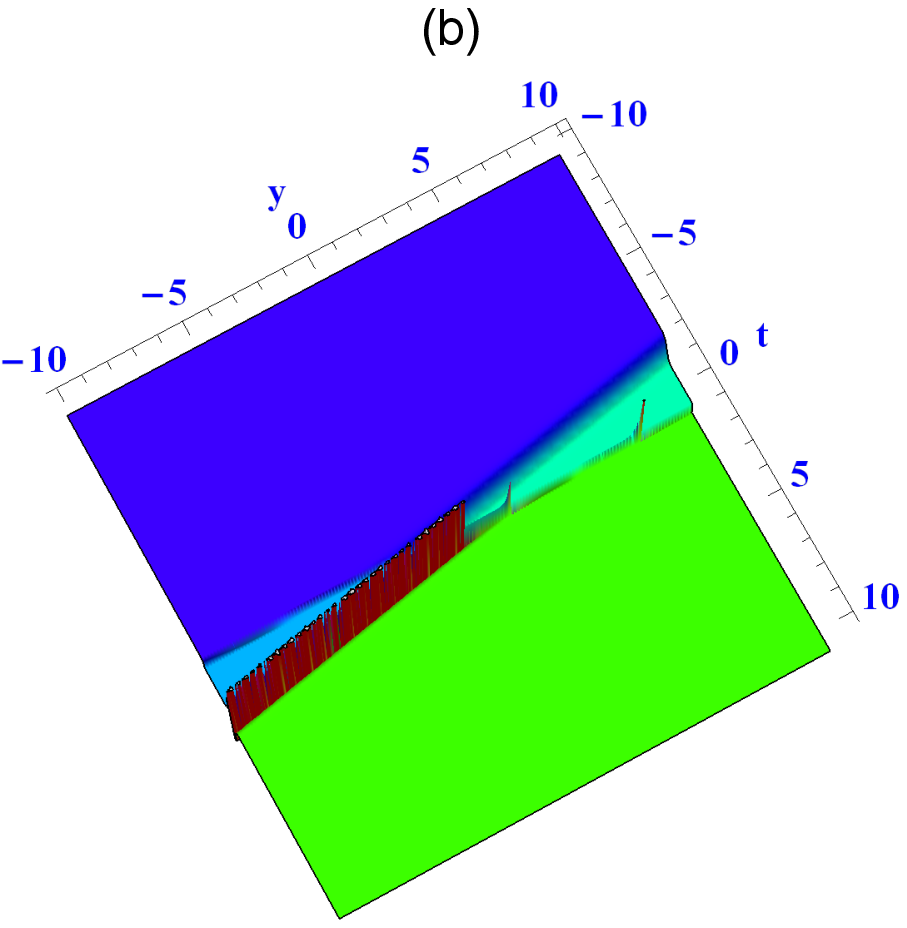}
\includegraphics[scale=0.4,bb=-260 270 10 10]{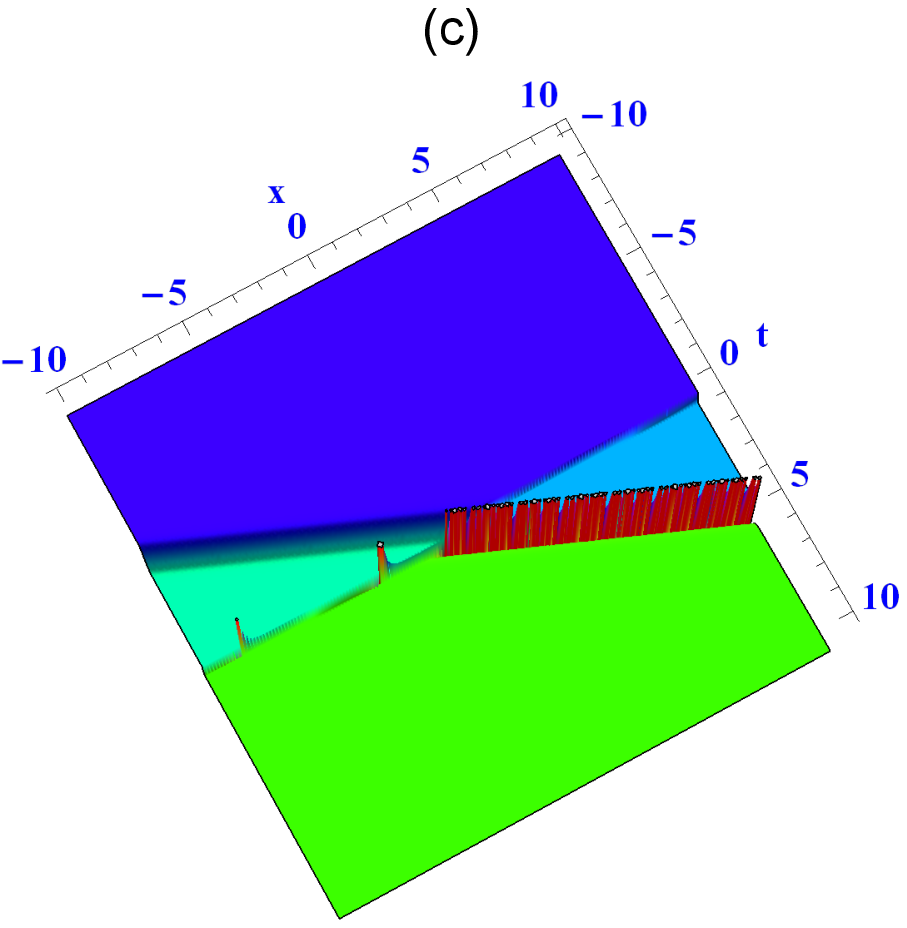}
\vspace{4cm}
\begin{tabbing}
\textbf{Fig. 1}. Multi-wave solution (6) with  (a) $t=0$, (b)
$x=0$, (c) $y=0$.\\
\end{tabbing}

\quad Secondly, we select
\begin{eqnarray}a(t)&=&\chi_4=1,
c(t)=t^2, \varphi_1=3, \varphi_2=\varphi_5=-2,\nonumber\\
\chi_3&=&\chi_5=\varphi_4=2, \chi_1=4,
 \epsilon_1=1,  \varphi_7=-1.\nonumber
\end{eqnarray}Then, the corresponding multi-wave solution can be
read as
\begin{eqnarray}u&=&[2 [-8 \sin [\frac{1}{40} \left(\frac{80 t^3}{3}-3376 t\right)+2 x-2 y+1]-3
   \sqrt{\frac{13}{5}} \sinh [\frac{1}{520} (\frac{80 t^3}{3}\nonumber\\&-&44176 t)-x-\frac{2
   y}{13}+2]+3 \cosh [\frac{1}{40} \left(\frac{80 t^3}{3}-3856 t\right)+3 x-2
   y+2]]]\nonumber\\&/&[4 \cos [\frac{1}{40} \left(\frac{80 t^3}{3}-3376 t\right)+2 x-2
   y+1]+\sinh [\frac{1}{40} \left(\frac{80 t^3}{3}-3856 t\right)\nonumber\\&+&3 x-2 y+2]+3
   \sqrt{\frac{13}{5}} \cosh [\frac{1}{520} \left(\frac{80 t^3}{3}-44176 t\right)-x-\frac{2
   y}{13}+2]].
\end{eqnarray}Fig. 2(a), Fig. 2(b) and Fig. 2(c) describe the propagation characteristics of Eq. (7) at  $t = 0$,
$x = 0$ and $y = 0$, respectively. It mainly shows the influence of variable coefficients on the multi-wave solution (5).

\includegraphics[scale=0.4,bb=20 270 10 10]{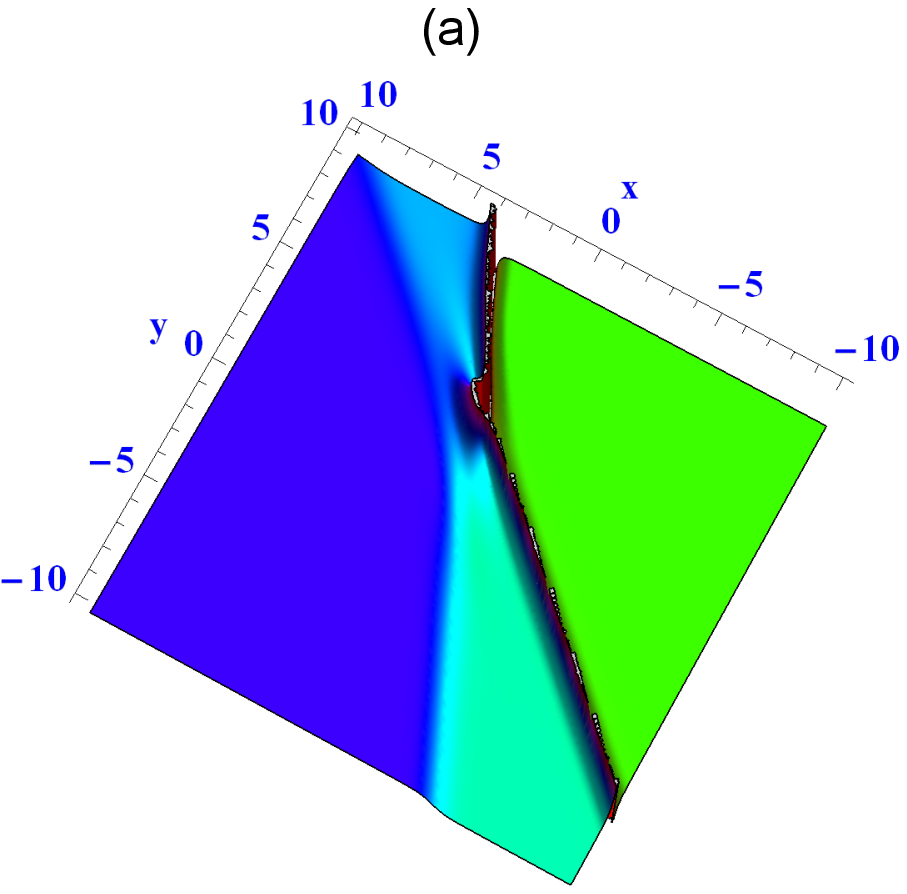}
\includegraphics[scale=0.4,bb=-255 270 10 10]{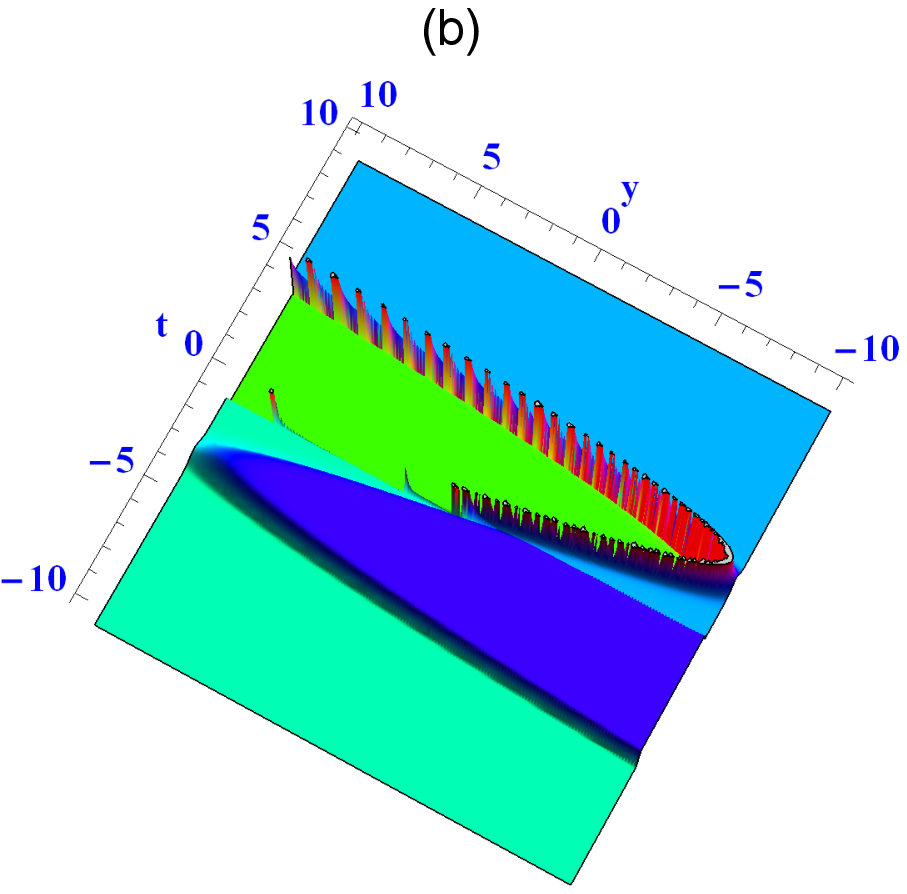}
\includegraphics[scale=0.4,bb=-260 270 10 10]{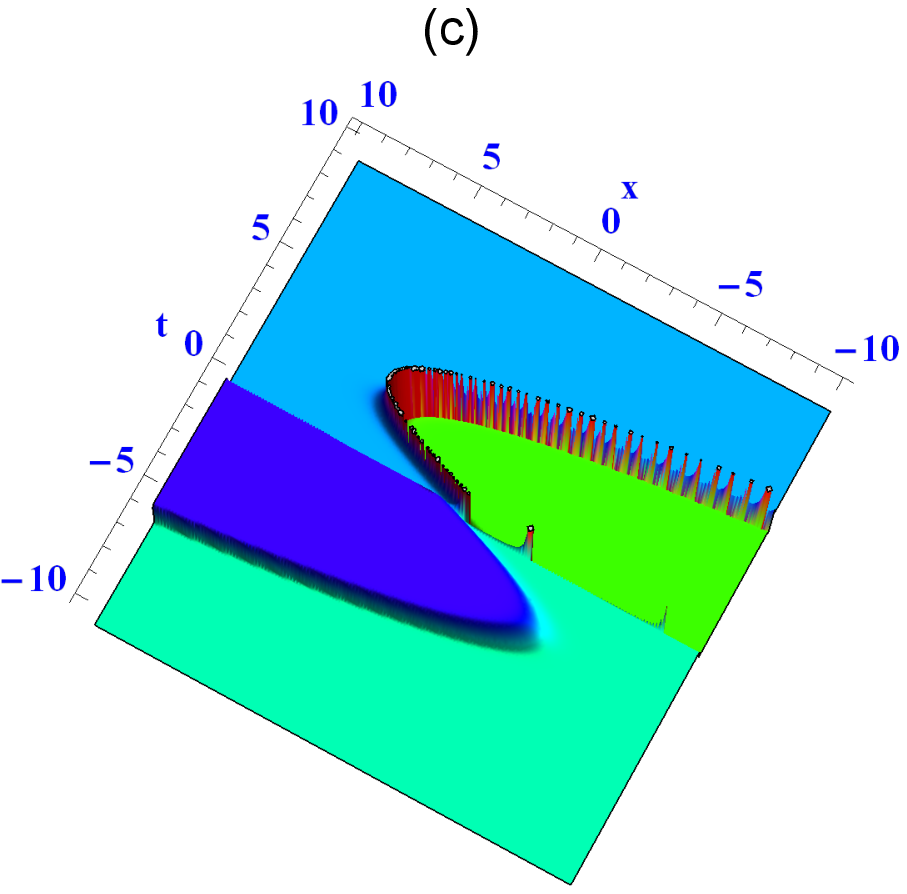}
\vspace{4cm}
\begin{tabbing}
\textbf{Fig. 2}. Multi-wave solution (7) with  (a) $t=0$, (b) $x=0$,
(c) $y=0$.
\end{tabbing}

\section{Breather wave solutions}
\label{sec:3} \quad Breathers are special solitons with periodic structure localized in space. It is often used to explain the generation of rogue waves and the nonlinear stage of modulation instability. They are mainly divided into Kuznetso-Ma breathers, Akhmediev breathers and general breathers. When the period of Kuznetso-Ma and Akhmediev breathers approaches infinity, we can get a peregrine soliton, which can help us understand the dynamic properties of rogue waves [36-38]. Based on the homoclinic breather approach [39],
we look for the breather wave solutions of Eq. (1) by using the
following assumption {\begin{eqnarray} \xi&=& \theta _2(t) \sin
[\varphi _6(t)+\varphi _4 x+\varphi _5 y]+\theta _3(t) \cos [\varphi
_9(t)+\varphi _7 x+\varphi _8 y]\nonumber\\&+&\theta _1(t)
e^{\varphi _3(t)+\varphi _1 x+\varphi _2 y}+e^{-\varphi
   _3(t)-\varphi _1 x-\varphi _2 y},
\end{eqnarray}}where  $\varphi_i(1\leq i \leq 9)$ and $\theta_i(t) (i=1,2,3)$  are
unknown parameters. Substituting Eq. (8) into Eq. (3), the breather
wave solutions of Eq. (1) are derived as follows
\begin{eqnarray}u&=&[2 [-\varphi _7 \chi _9 \sqrt{\theta _1(t)} \sin [\varphi _9(t)+\varphi _7 x+\varphi _8 y]+\varphi _4 \chi _{10} \sqrt{\theta _1(t)} \cos
   [\varphi _6(t)+\varphi _4 x\nonumber\\&+&\varphi _5 y]+\varphi _1 \theta _1(t) e^{\varphi _3(t)+\varphi _1 x+\varphi _2 y}+\varphi _1 \left(-e^{-\varphi
   _3(t)-\varphi _1 x-\varphi _2 y}\right)]]\nonumber\\&/&[\chi _{10} \sqrt{\theta _1(t)} \sin [\varphi _6(t)+\varphi _4 x+\varphi _5 y]+\chi _9 \sqrt{\theta
   _1(t)} \cos [\varphi _9(t)+\varphi _7 x+\varphi _8 y]\nonumber\\&+&\theta _1(t) e^{\varphi _3(t)+\varphi _1 x+\varphi _2 y}+e^{-\varphi _3(t)-\varphi _1
   x-\varphi _2 y}].
\end{eqnarray}All parameters have been interpreted in Appendix B.

\quad  In order to analyze the physical structure of the breather
wave solutions (9), two illustrated examples are presented because
of the existence of $\sin$ and $\cos$ in Eq. (8).

\quad First, we assume
\begin{eqnarray}b(t)&=&c(t)=\varphi_2=\varphi_5=1, \varphi_1=-1, \varphi_7=3,
\varphi_4=-2,\nonumber\\ \chi_6&=&\chi_7=\chi_8=\chi_9=0,
 \epsilon_2=1,\nonumber.
\end{eqnarray}Then, the corresponding breather
wave solution can be written as
\begin{eqnarray}u&=&\frac{2 [e^{\frac{3400 t}{27}+x-y}-e^{-\frac{3400 t}{27}-x+y}-4 \sqrt{2} \cos \left(\frac{1150 t}{27}+2 x-y\right)]}{e^{\frac{3400
   t}{27}+x-y}+e^{-\frac{3400 t}{27}-x+y}-2 \sqrt{2} \sin \left(\frac{1150 t}{27}+2 x-y\right)}.
\end{eqnarray}Fig. 3(a), Fig. 3(b) and Fig. 3(c) represent the propagation characteristics of Eq. (10) at $t = 0$,
$x = 0$ and $y = 0$, respectively.  Akhmediev breathers can be seen in Fig. 3.

\includegraphics[scale=0.4,bb=20 270 10 10]{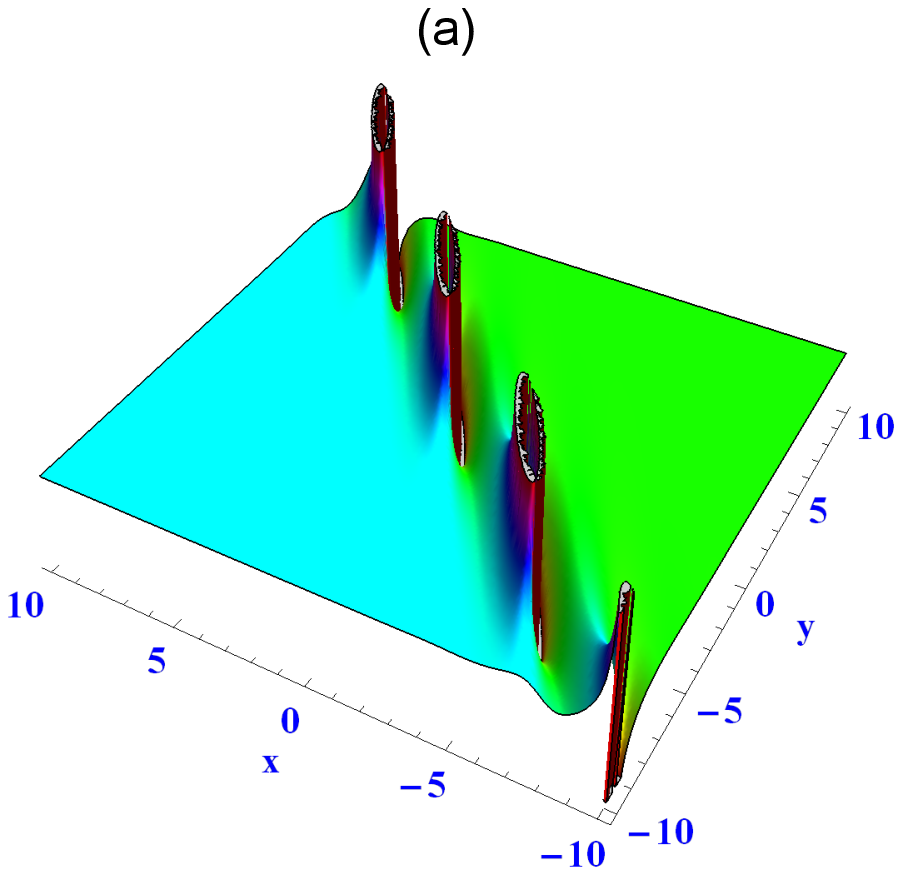}
\includegraphics[scale=0.4,bb=-255 270 10 10]{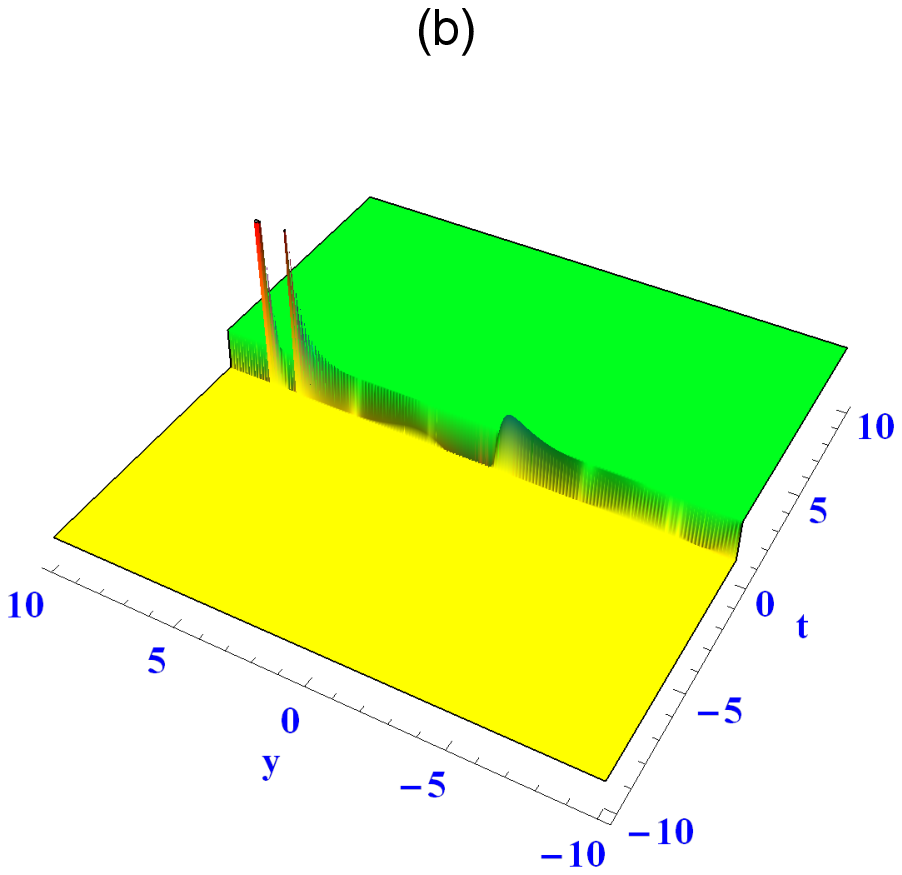}
\includegraphics[scale=0.4,bb=-260 270 10 10]{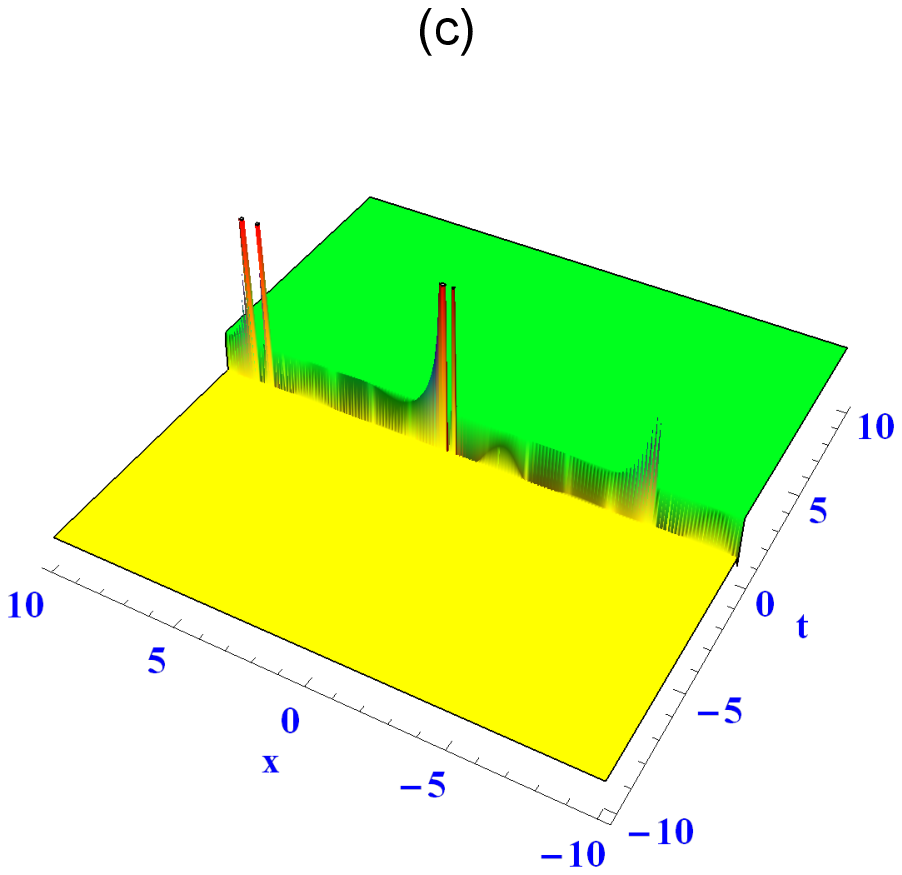}
\vspace{4cm}
\begin{tabbing}
\textbf{Fig. 3}. breather wave solution (10) with
 (a) $t=0$, (b) $x=0$, (c) $y=0$.\\
\end{tabbing}

\quad Secondly, we select
\begin{eqnarray}b(t)&=&c(t)=\varphi_2=\varphi_5=1, \varphi_1=-1, \varphi_7=3,
\varphi_4=-2,\nonumber\\ \chi_6&=&\chi_7=\chi_8=0, \chi_9=4,
 \epsilon_2=1.\nonumber
\end{eqnarray}Then, the corresponding breather wave solution can be
read as
\begin{eqnarray}u&=&[2 [e^{\frac{3400 t}{27}+x-y}-e^{-\frac{3400 t}{27}-x+y}-12 \sin \left(\frac{9350 t}{81}+3 x-\frac{5 y}{3}\right)\nonumber\\&-&
4 \sqrt{10} \cos \left(\frac{1150
   t}{27}+2 x-y\right)]]/[e^{\frac{3400 t}{27}+x-y}+e^{-\frac{3400 t}{27}-x+y}\nonumber\\&-&2 \sqrt{10} \sin \left(\frac{1150 t}{27}+2 x-y\right)
   +4 \cos \left(\frac{9350
   t}{81}+3 x-\frac{5 y}{3}\right)].
\end{eqnarray}Fig. 4(a), Fig. 4(b) and Fig. 4(c) show the propagation characteristics of Eq. (11) at time $t = 0$,
$x = 0$ and $y = 0$, respectively.  Double Akhmediev breathers can be seen in Fig. 4.

\includegraphics[scale=0.4,bb=20 270 10 10]{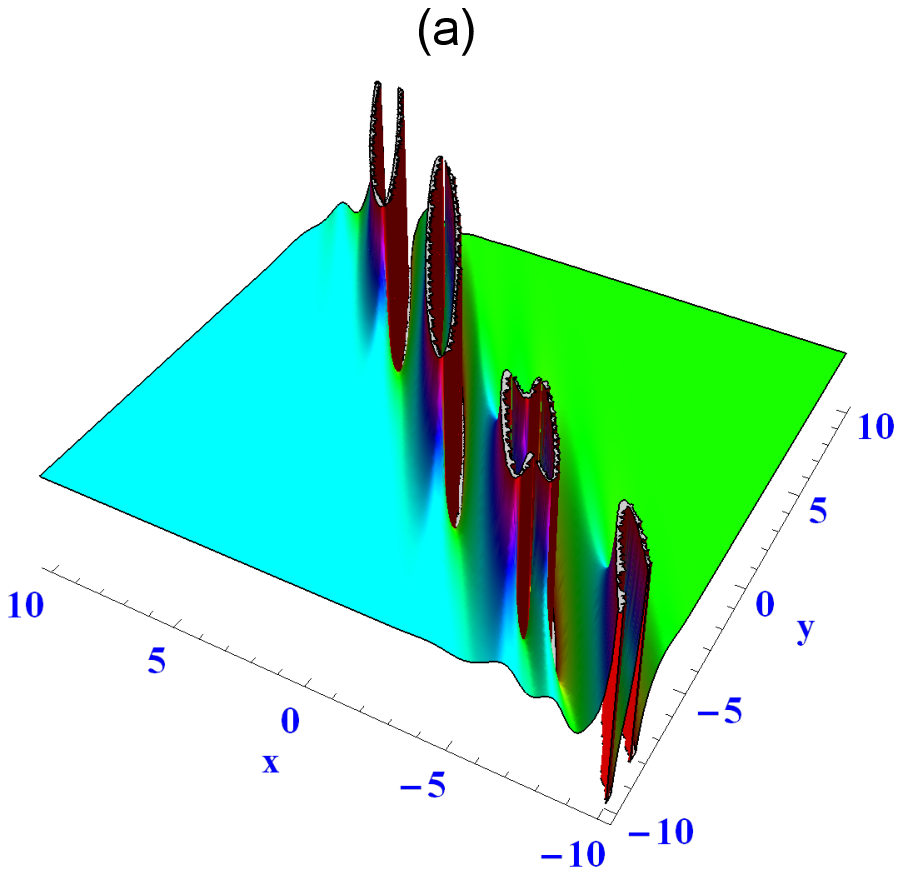}
\includegraphics[scale=0.4,bb=-255 270 10 10]{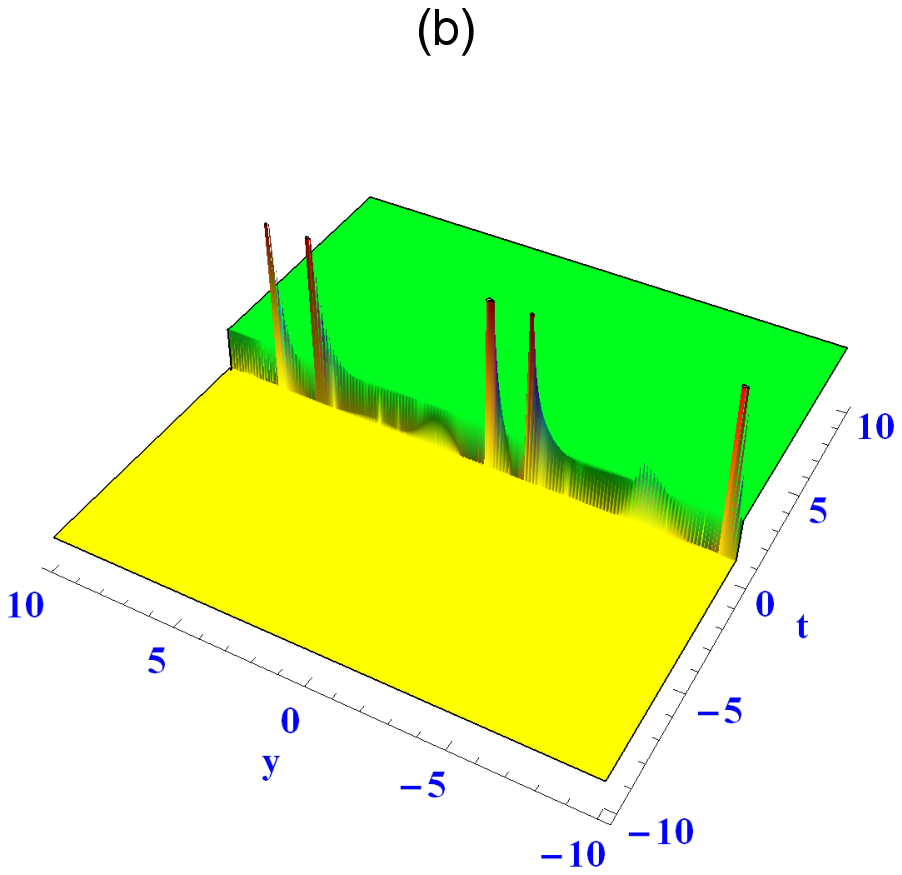}
\includegraphics[scale=0.4,bb=-260 270 10 10]{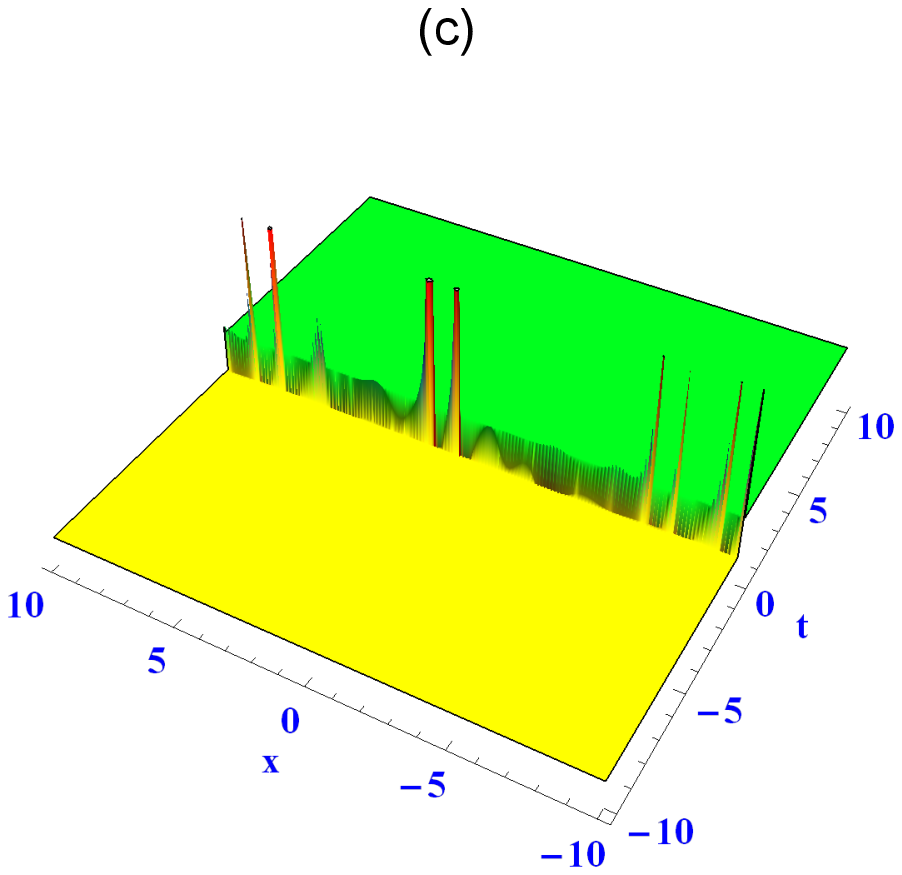}
\vspace{4cm}
\begin{tabbing}
\textbf{Fig. 4}. Breather wave solution (11) with
 (a) $t=0$, (b) $x=0$, (c) $y=0$.
\end{tabbing}

\section{Lump solutions}
\label{sec:3} \quad Rogue waves are localized in both space and time. Nowadays, people are more and more interested in the rational solitons which are locally confined to space, which is called the lump wave [40]. Based on the Hirota's bilinear method with
variable coefficients [41,42], lump solution of Eq. (1) are supposed
as{\begin{eqnarray} \xi&=&\alpha _7(t)+[\int \alpha _3(t) dt+\alpha
_1 x+\alpha _2 y]{}^2+[\int \alpha _6(t) dt+\alpha _4
   x+\alpha _5 y]{}^2,
\end{eqnarray}}where $\alpha_i(1\leq i \leq 7)$ is
undetermined parameter. Substituting Eq. (12) into Eq. (3), the
corresponding lump solution is
 obtained as follows
\begin{eqnarray}u&=&[4 [-\alpha _2 \alpha _1 [\int \frac{\alpha _2 [\alpha _1 (b(t) c(t)-a(t)
   k(t))+\alpha _2 b(t) k(t)]+\alpha _4^2 a(t) c(t)}{\alpha _4^2 a(t)^2+\alpha _2^2 b(t)^2} \,
   dt]\nonumber\\&+&\alpha _4 [\int \frac{\alpha _4^2 \alpha _2 \left(\alpha _1 a(t) c(t)+\alpha _2
   a(t) k(t)-\alpha _2 b(t) c(t)\right)+\alpha _1 \alpha _2^3 b(t) k(t)}{\alpha _4^3 a(t)^2+\alpha
   _2^2 \alpha _4 b(t)^2} \, dt\nonumber\\&+&\alpha _4 x]+\alpha _1^2 x]]/[[\alpha _2
   [y-\int [\alpha _2 [\alpha _1 [b(t) c(t)-a(t) k(t)]+\alpha _2 b(t)
   k(t)]\nonumber\\&+&\alpha _4^2 a(t) c(t)]/[\alpha _4^2 a(t)^2+\alpha _2^2 b(t)^2] \, dt]+\alpha _1
   x]{}^2+[\alpha _4 x-\frac{\alpha _1 \alpha _2 y}{\alpha
   _4}\nonumber\\&+&\int \frac{\alpha _4^2 \alpha _2 [a(t) \left(\alpha _1 c(t)+\alpha _2
   k(t)\right)-\alpha _2 b(t) c(t)]+\alpha _1 \alpha _2^3 b(t) k(t)}{\alpha _4^3 a(t)^2+\alpha
   _2^2 \alpha _4 b(t)^2} \, dt]{}^2].
\end{eqnarray}Other four different lump solutions are presented in Appendix
C.

\quad  In order to analyze the physical structure of the lump
solution (13), three illustrated examples are listed  thanks to the
existence of variable coefficients in Eq. (1).

\quad First, we suppose
\begin{eqnarray} a(t)=c(t)=b(t)=1, k(t)=-3,
 \alpha_2=-2, \alpha_4=-1, \alpha_1=2.\nonumber
\end{eqnarray}Then, the corresponding lump solution can be written as
\begin{eqnarray}u=\frac{4 (5 x-16 t)}{\left(-\frac{28 t}{5}-x-4 y\right)^2+\left(2 x-2 \left(\frac{27
   t}{5}+y\right)\right)^2}.
\end{eqnarray}Fig. 5(a), Fig. 5(b) and Fig. 5(c) describe the propagation characteristics of Eq. (14) at  $t = 0$,
$x = 0$ and $y = 0$, respectively.  In Fig. 5,  the bright-dark lump wave can be found, which has one peak and one valley. Their peak and valley are meristic.

\includegraphics[scale=0.4,bb=20 270 10 10]{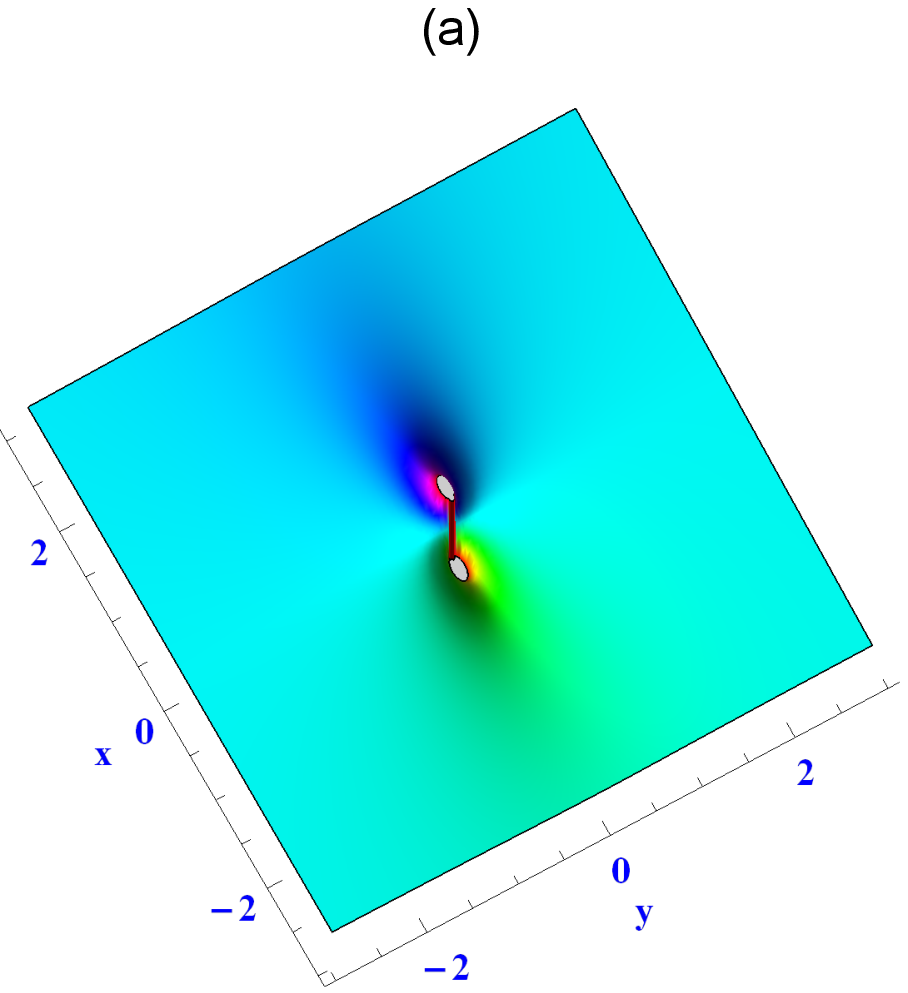}
\includegraphics[scale=0.4,bb=-255 270 10 10]{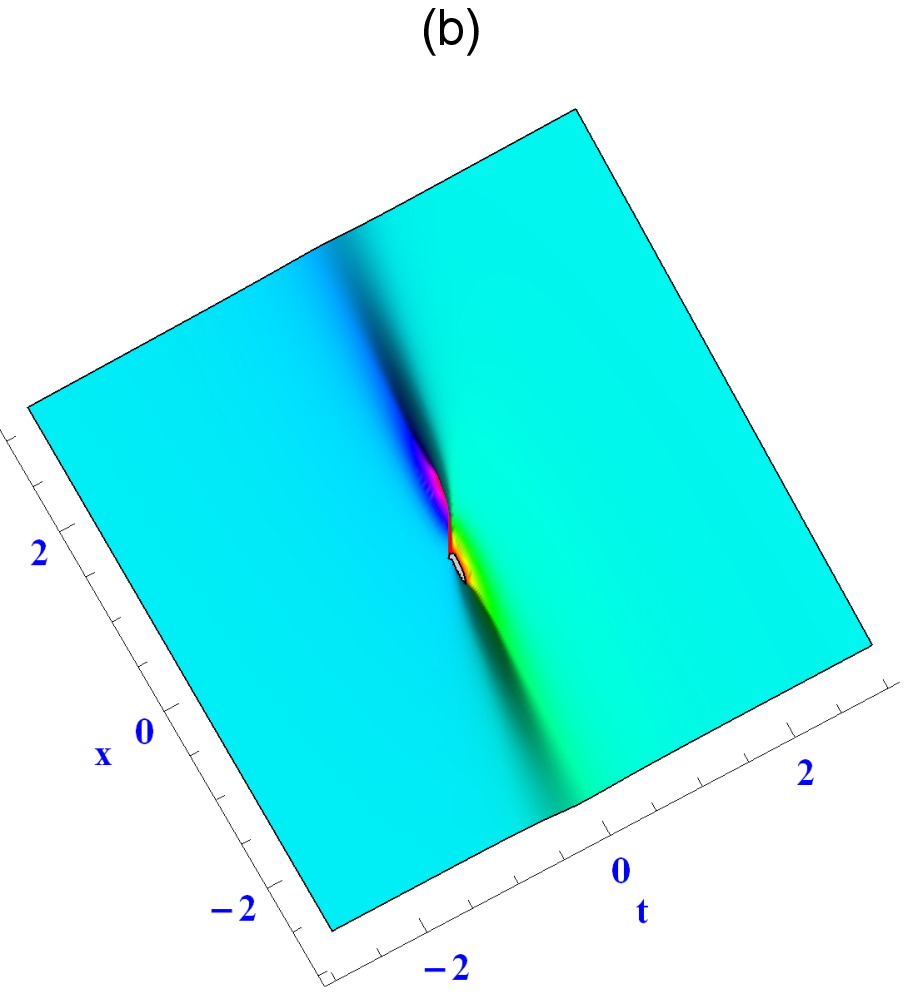}
\includegraphics[scale=0.4,bb=-260 270 10 10]{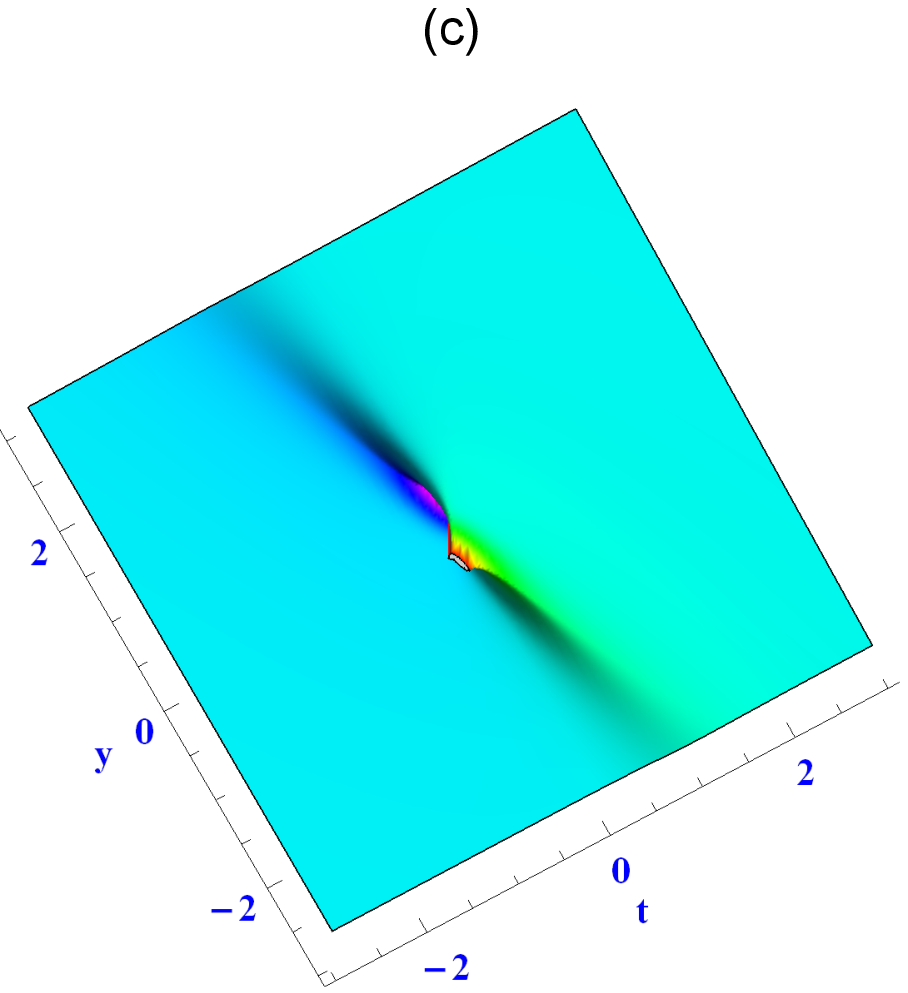}
\vspace{4.1cm}
\begin{tabbing}
\textbf{Fig. 5}. Lump solution (14) with (a) $t=0$, (b) $x=0$, (c)
$y=0$.\\
\end{tabbing}

\quad Secondly, we select
\begin{eqnarray}a(t)=b(t)=1, c(t)=t, k(t)=2 t,
 \alpha_2=3, \alpha_4=-3, \alpha_1=2.\nonumber
\end{eqnarray}Then, the corresponding lump solution can be
read as
\begin{eqnarray}u=\frac{4 [-3 \left(-\frac{9 t^2}{4}-3 x\right)-\frac{7 t^2}{2}+4 x]}{\left(-\frac{9
   t^2}{4}-3 x+2 y\right)^2+[3 \left(y-\frac{7 t^2}{12}\right)+2 x]^2}.
\end{eqnarray}Fig. 6(a), Fig. 6(b) and Fig. 6(c) show the propagation characteristics of Eq. (15) at $t = 0$,
$x = 0$ and $y = 0$, respectively. Fig. 6 mainly demonstrates the influence of variable coefficients on the bright-dark lump wave.

\includegraphics[scale=0.4,bb=20 280 10 10]{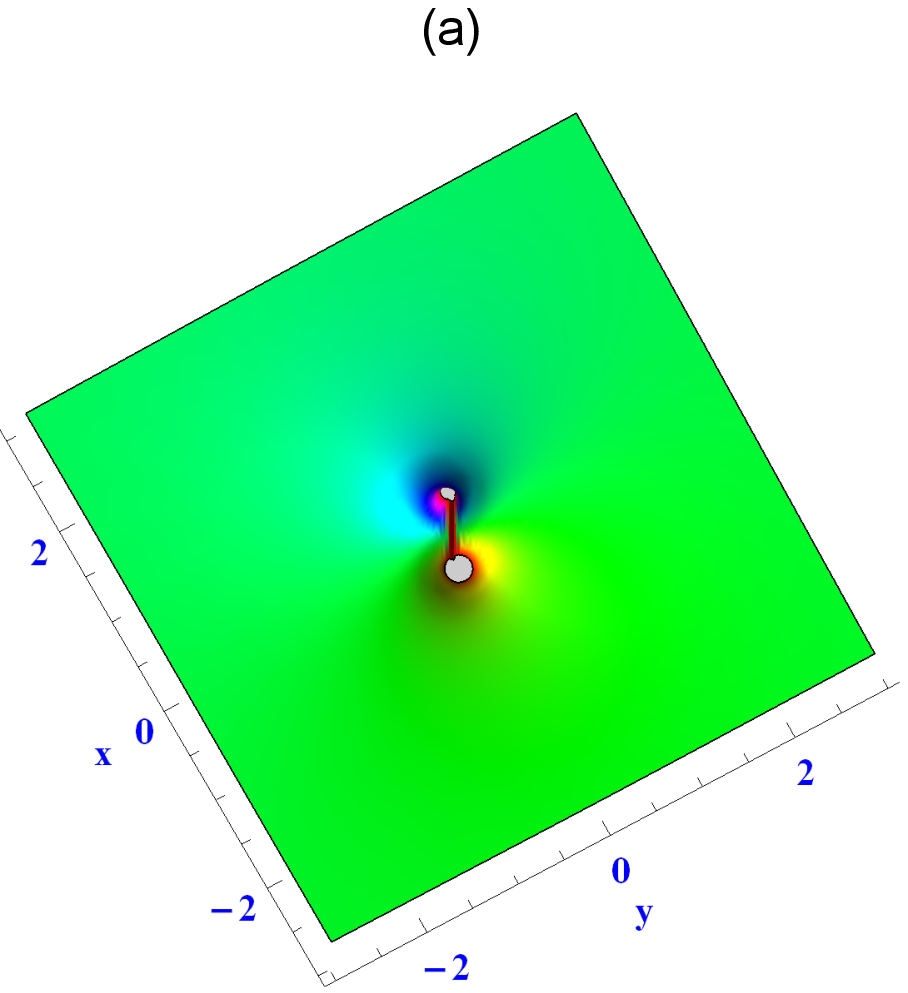}
\includegraphics[scale=0.4,bb=-255 280 10 10]{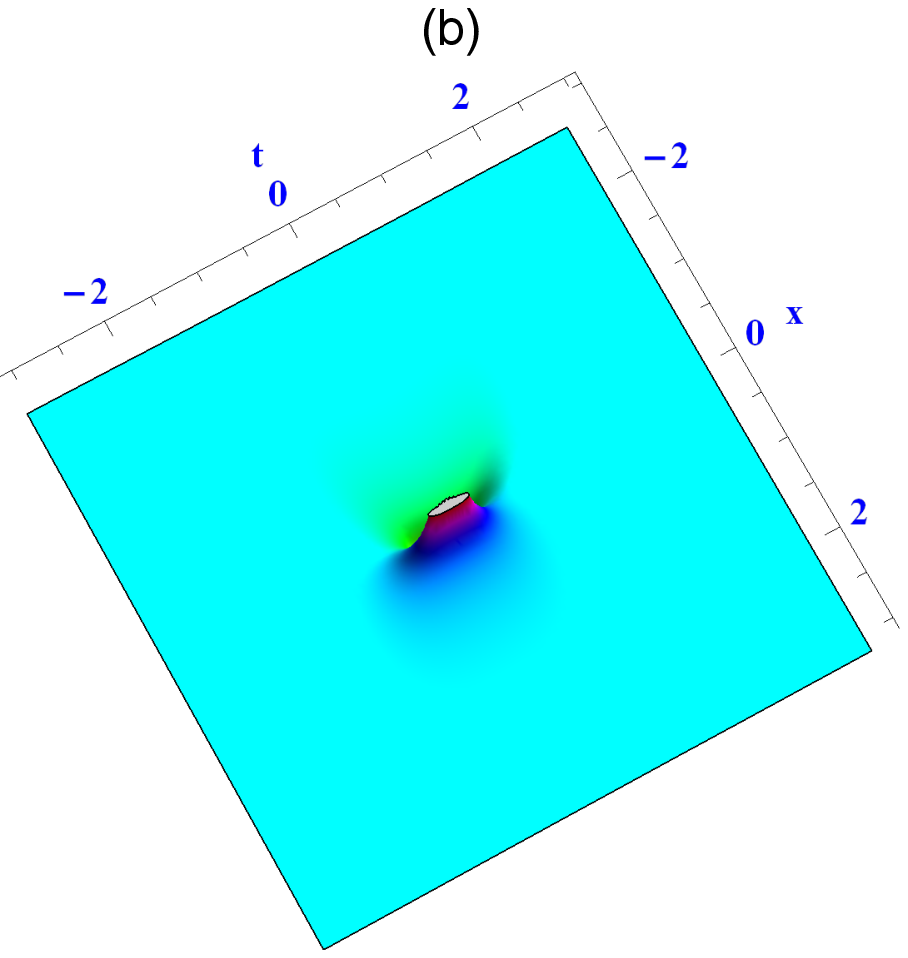}
\includegraphics[scale=0.4,bb=-260 280 10 10]{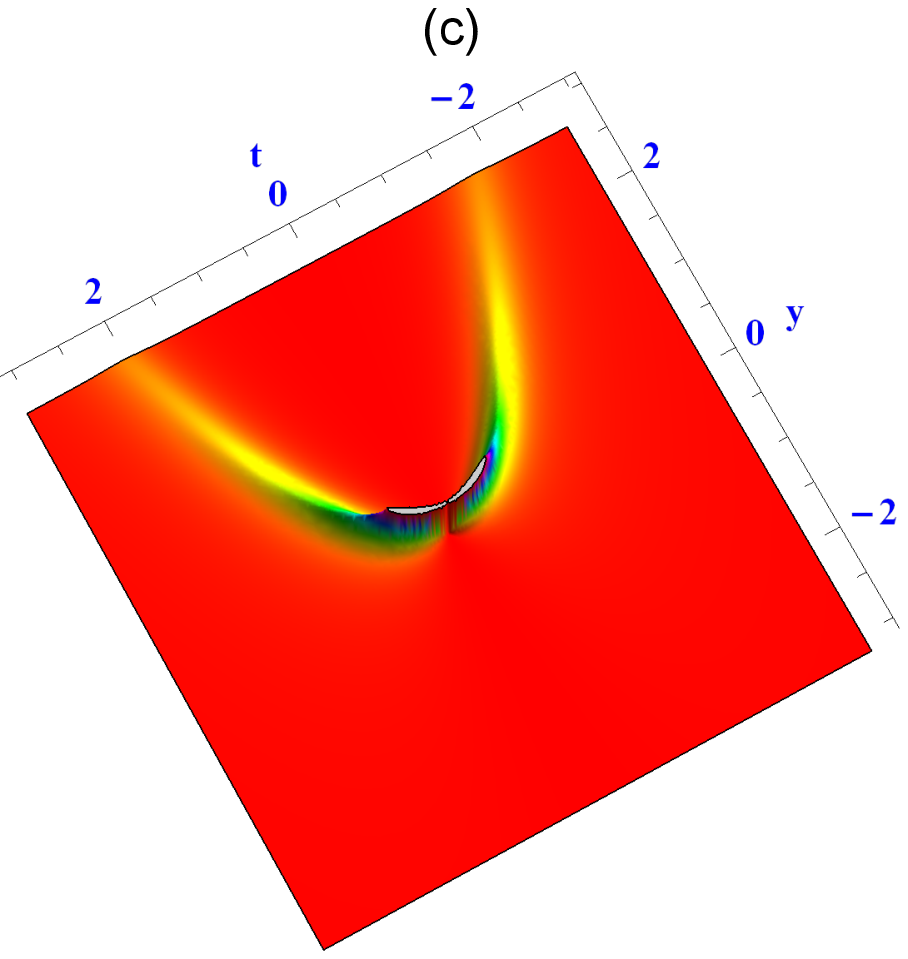}
\vspace{4.1cm}
\begin{tabbing}
\textbf{Fig. 6}. Lump solution (15) with  (a) $t=0$, (b) $x=0$, (c)
$y=0$.\\
\end{tabbing}

\quad Third, we choose
\begin{eqnarray}a(t)=b(t)=1, c(t)=\sin t, k(t)=\cos t,
 \alpha_2=3, \alpha_4=-3, \alpha_1=2.\nonumber
\end{eqnarray}Then, the corresponding lump solution can be
read as
\begin{eqnarray}u=\frac{4 [-3 [-\frac{5 \sin (t)}{2}-\frac{\cos (t)}{2}-3 x]-6 [\frac{\sin
   (t)}{6}-\frac{5 \cos (t)}{6}]+4 x]}{[3 [-\frac{\sin (t)}{6}+\frac{5 \cos
   (t)}{6}+y]+2 x]^2+[-\frac{5 \sin (t)}{2}-\frac{\cos (t)}{2}-3 x+2 y]^2}.
\end{eqnarray}Fig. 7(a), Fig. 7(b) and Fig. 7(c) show the propagation characteristics of Eq. (16) at time $t = 0$,
$x = 0$ and $y = 0$, respectively.In Fig. 7, we can observe periodic-type lump waves. They all appear in pairs, interact, and spread forward.

\includegraphics[scale=0.4,bb=20 270 10 10]{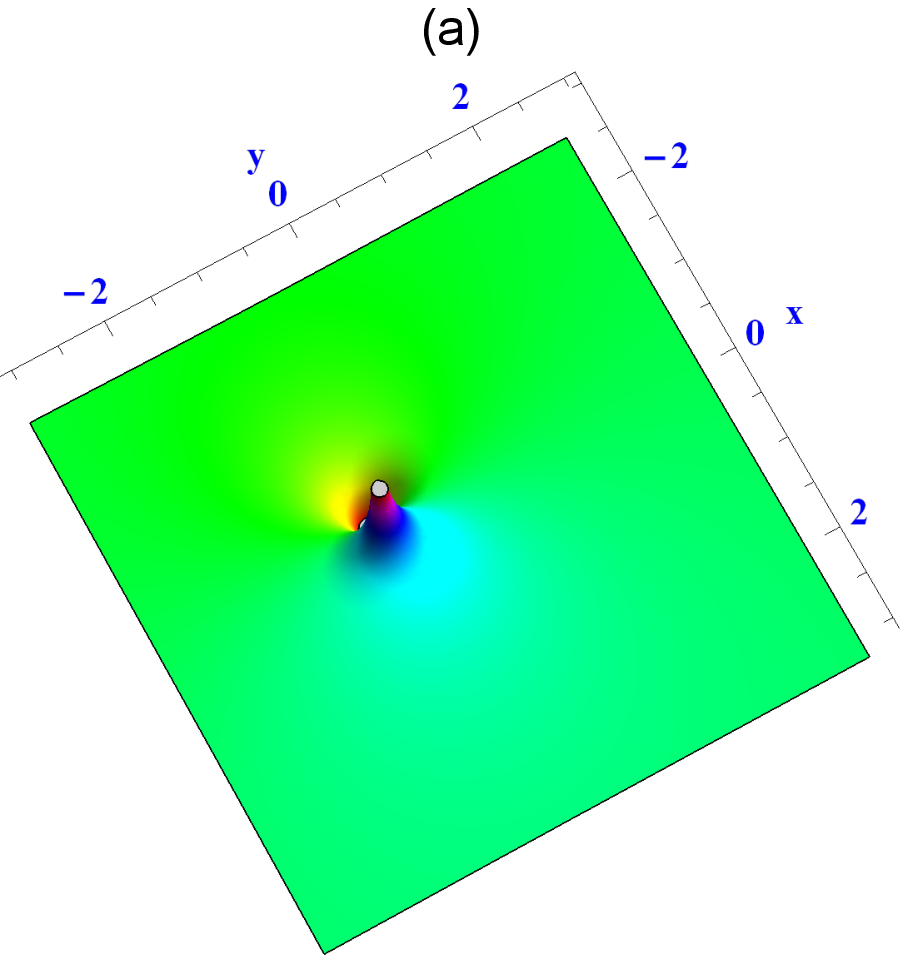}
\includegraphics[scale=0.4,bb=-255 270 10 10]{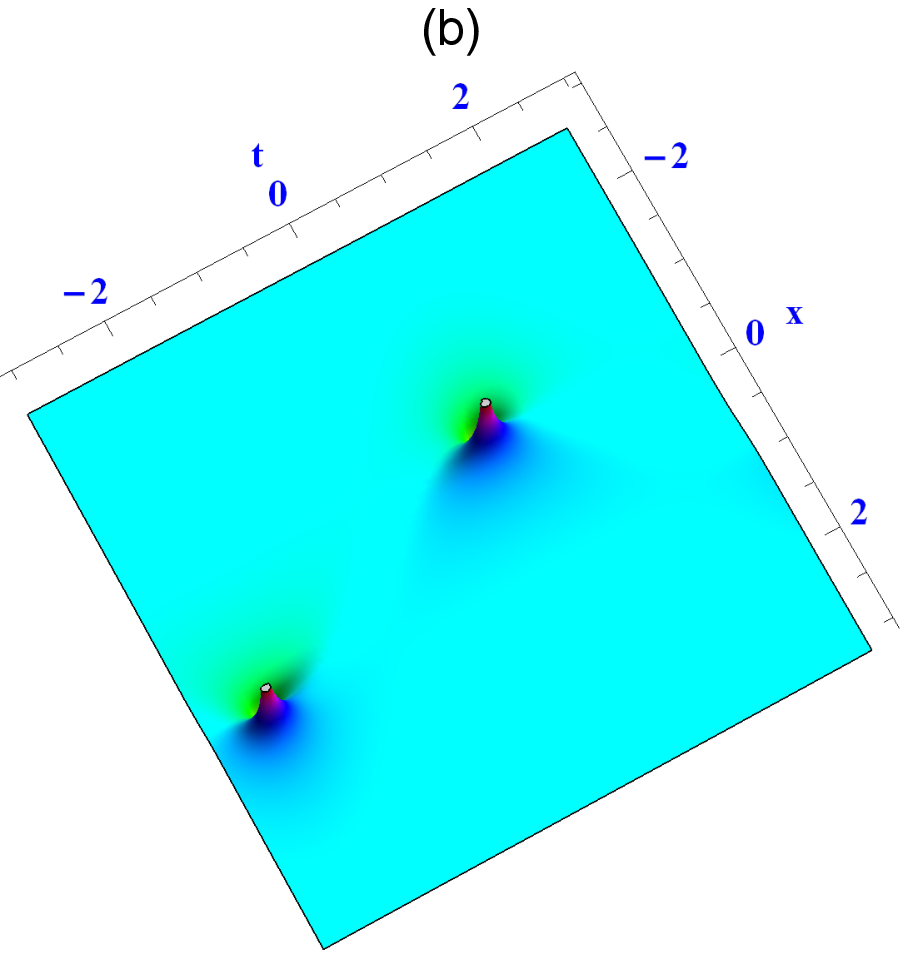}
\includegraphics[scale=0.4,bb=-260 270 10 10]{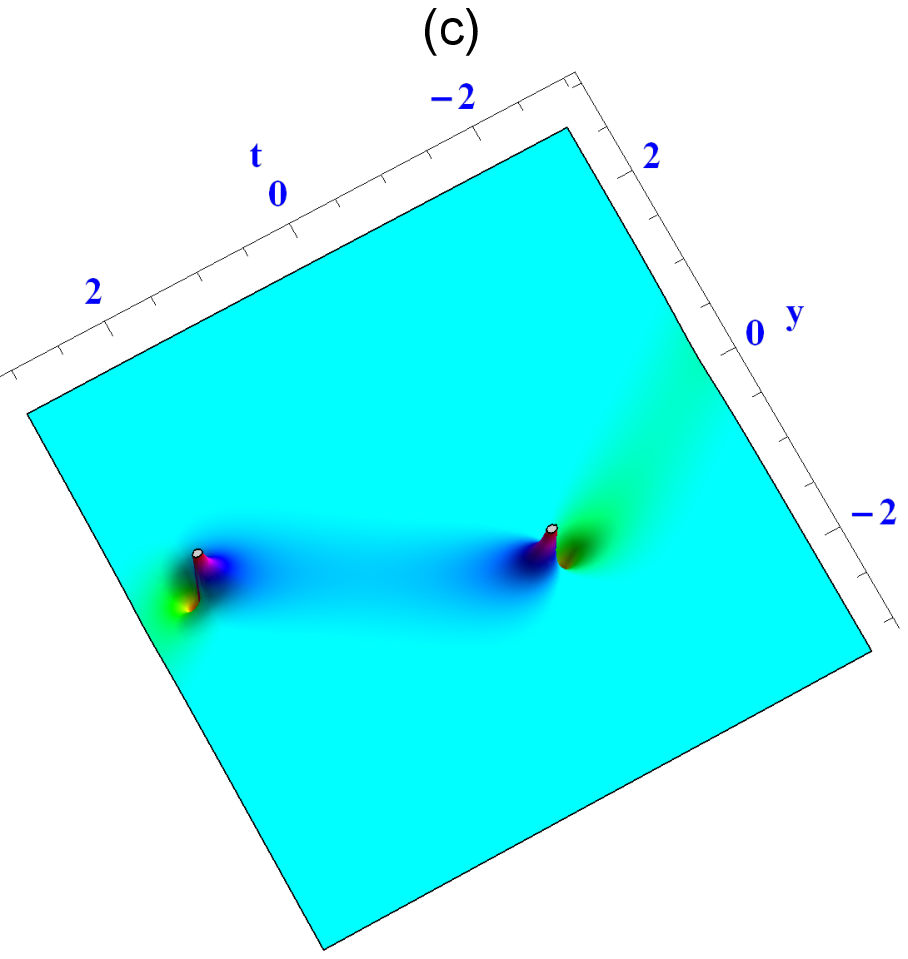}
\vspace{4.1cm}
\begin{tabbing}
\textbf{Fig. 7}. Lump solution (16) with  (a) $t=0$, (b) $x=0$, (c)
$y=0$.
\end{tabbing}

\section{ Conclusion}
\label{sec:3} \quad In this paper, the (2+1)-dimensional vcBLMPe is
studied. Multi-wave, breather wave and lump solutions are presented
by using the three waves method, the homoclinic breather approach
and the Hirota's bilinear method with variable coefficients. All
results have not been studied in previous literature. Fig. 1
describes the physical structure of multi-wave solution (6) when
$c(t)$ is a constant. Fig. 2 shows the physical structure of
multi-wave solution (7)  when $c(t)$ is a function. Fig. 3 and Fig.4
show the propagation characteristics of  breather wave solution (9)
and (10) when $\chi_9$ selects different value. Figs. 5-7 represent
a lump wave when $c(t)$ and $k(t)$ choose different functions.\\

\noindent {\bf Compliance with ethical standards}\\

\quad {\bf Conflict of interests} The authors declare that there is
no conflict of interests regarding the publication of this article.

\quad {\bf Ethical standard} The authors state that this research
complies with ethical standards. This research does not involve
either human participants or animals.



\noindent {\bf Appendix A}\\ \begin{eqnarray} \theta_2(t)&=&\chi _1
\theta_1(t), \theta_3(t)=\chi _2 \theta_1(t),
b(t)=-\frac{\left(\varphi _1^2+\varphi _4^2\right) a(t)}{3
\left(\varphi _1 \varphi _2+\varphi _4
   \varphi _5\right)},\nonumber\\
 \varphi_8&=&\frac{-\varphi _4 \varphi _5 \varphi _1^2+\varphi _2 \left(\varphi _4^2+\varphi _7^2\right) \varphi
   _1+\varphi _4 \varphi _5 \varphi _7^2}{\left(\varphi _1^2+\varphi _4^2\right) \varphi _7},
 \nonumber\\
\varphi_9(t)&=& [\int [-\left(\varphi _2 \varphi _4-\varphi _1
\varphi _5\right){}^2 \left(\varphi _1
   \varphi _2+\varphi _4 \varphi _5\right) [-\varphi _4 \varphi _5 \varphi _1^2+\varphi _2
   \left(\varphi _4^2+\varphi _7^2\right) \varphi _1\nonumber\\&+&\varphi _4 \varphi _5 \varphi
   _7^2]
   c(t)+\left(\varphi _1 \varphi _2+\varphi _4 \varphi _5\right){}^2 [\left(2 \varphi
   _2^2-\varphi _5^2\right) \varphi _1^2+6 \varphi _2 \varphi _4 \varphi _5 \varphi _1\nonumber\\&-&\varphi _4^2
   \left(\varphi _2^2-2 \varphi _5^2\right)] \varphi _7^4+\left(\varphi _2 \varphi _4-\varphi
   _1 \varphi _5\right) \left(\varphi _1 \varphi _2+\varphi _4 \varphi _5\right) [\varphi _4
   \left(\varphi _1^3-\varphi _1 \varphi _4^2\right) \varphi _2^2\nonumber\\&+&4 \left(\varphi _1^4+3 \varphi
   _4^2 \varphi _1^2+\varphi _4^4\right) \varphi _5 \varphi _2+\varphi _1 \varphi _4 \left(\varphi
   _4^2-\varphi _1^2\right) \varphi _5^2] \varphi _7^2\nonumber\\&+&\varphi _1 \varphi _4 \left(\varphi _1
   \varphi _5-\varphi _2 \varphi _4\right){}^3 \left(\varphi _1^3 \varphi _2-\varphi _4^3 \varphi
   _5\right)]/a(t) \, dt]\nonumber\\&/&[\left(\varphi _1^2+\varphi _4^2\right) \left(\varphi _2 \varphi
   _4-\varphi _1 \varphi _5\right){}^2 \left(\varphi _1 \varphi _2+\varphi _4 \varphi _5\right)
   \varphi _7]+\chi _3,\nonumber\\
   \varphi_6(t)&=&[\int [\left(\varphi _1 \varphi _2+\varphi _4 \varphi _5\right) [\left(5 \varphi
   _2^2 \varphi _5-\varphi _5^3\right) \varphi _1^2-3 \varphi _2 \varphi _4 \left(\varphi _2^2-3
   \varphi _5^2\right) \varphi _1\nonumber\\&+&2 \varphi _4^2 \varphi _5 \left(\varphi _5^2-2 \varphi
   _2^2\right)] \varphi _7^2-\left(\varphi _2 \varphi _4-\varphi _1 \varphi _5\right){}^2
   [\varphi _1 \varphi _5 \varphi _2 [c(t)+\varphi _1^2\nonumber\\&+&3 \varphi _4^2]+\varphi _4
   \varphi _5^2 \left(c(t)-\varphi _4^2\right)+3 \varphi _1^2 \varphi _4 \varphi _2^2]]\nonumber\\&/&a(t)
   \, dt]/[\left(\varphi _2 \varphi _4-\varphi _1 \varphi _5\right){}^2 \left(\varphi _1 \varphi
   _2+\varphi _4 \varphi _5\right)]+\chi
   _4,\nonumber\\
\varphi_3(t)&=&[\int [\left(\varphi _1 \varphi _2+\varphi _4 \varphi
_5\right) [2 \left(\varphi
   _2^3-2 \varphi _2 \varphi _5^2\right) \varphi _1^2-3 \varphi _4 \varphi _5 \left(\varphi _5^2-3
   \varphi _2^2\right) \varphi _1\nonumber\\&-&\varphi _2 \varphi _4^2 \left(\varphi _2^2-5 \varphi
   _5^2\right)] \varphi _7^2-\left(\varphi _2 \varphi _4-\varphi _1 \varphi _5\right){}^2
   [\varphi _1 \varphi _2^2 \left(c(t)+\varphi _1^2\right)\nonumber\\&+&\varphi _4 \varphi _5 \varphi _2
   \left(c(t)-3 \varphi _1^2-\varphi _4^2\right)-3 \varphi _1 \varphi _4^2 \varphi
   _5^2]]/a(t) \, dt]\nonumber\\&/&[\left(\varphi _2 \varphi _4-\varphi _1 \varphi _5\right){}^2
   \left(\varphi _1 \varphi _2+\varphi _4 \varphi _5\right)]+\chi _5,\nonumber\\
   k(t)&=& -[\left(\varphi _1^2+\varphi _4^2\right) [\left(\varphi _2 \varphi _4-\varphi _1 \varphi
   _5\right){}^2 [\varphi _1 \varphi _2 \left(c(t)+\varphi _1^2\right)+\varphi _4 \varphi _5
   \left(c(t)-\varphi _4^2\right)]\nonumber\\&+&\left(\varphi _1 \varphi _2+\varphi _4 \varphi _5\right)
   [\left(4 \varphi _2^2+\varphi _5^2\right) \varphi _1^2+6 \varphi _2 \varphi _4 \varphi _5
   \varphi _1+\varphi _4^2 \left(\varphi _2^2+4 \varphi _5^2\right)] \varphi _7^2]]\nonumber\\&/&[3
   \left(\varphi _2 \varphi _4-\varphi _1 \varphi _5\right){}^2 \left(\varphi _1 \varphi _2+\varphi
   _4 \varphi _5\right){}^2],\nonumber\\ \chi_2&=&\epsilon_1 \sqrt{\varphi _1^2+\varphi _4^2} \sqrt{\frac{\chi _1^2}{\varphi _1^2-\varphi _7^2}-\frac{1}{\varphi
   _4^2+\varphi _7^2}},
\end{eqnarray}where $\chi _i (i=1,\cdots,5)$ is integral constant, $\epsilon_1=\pm1$.\\

\noindent {\bf Appendix B}\\ \begin{eqnarray}
k(t)&=&-[\left(\varphi _1^2+\varphi _4^2\right) [\left(\varphi _2
\varphi _4-\varphi _1 \varphi _5\right){}^2 [\varphi _1 \varphi _2
\left(c(t)+\varphi
   _1^2\right)+\varphi _4 \varphi _5 \left(c(t)-\varphi _4^2\right)]\nonumber\\&-&\left(\varphi _1 \varphi _2+\varphi _4 \varphi _5\right) [\left(4 \varphi
   _2^2+\varphi _5^2\right) \varphi _1^2+6 \varphi _2 \varphi _4 \varphi _5 \varphi _1+\varphi _4^2 \left(\varphi _2^2+4 \varphi _5^2\right)] \varphi
   _7^2]]\nonumber\\&/&[3 \left(\varphi _2 \varphi _4-\varphi _1 \varphi _5\right){}^2 \left(\varphi _1 \varphi _2+\varphi _4 \varphi _5\right){}^2], \nonumber\\
\varphi_3(t)&=&[\int [32 \varphi _2 \left(\varphi _1^2+\varphi
_4^2\right) \left(\varphi _2 \varphi _4-\varphi _1 \varphi
_5\right){}^2 \left(\varphi _1 \varphi
   _2+\varphi _4 \varphi _5\right) c(t)+32 \varphi _2^2 \varphi _5^2 \varphi _1^7\nonumber\\&-&32 \varphi _2 \varphi _4 \varphi _5 \left(2 \varphi _2^2+3 \varphi _5^2\right)
   \varphi _1^6+32 [\left(\varphi _2^4+7 \varphi _5^2 \varphi _2^2-3 \varphi _5^4\right) \varphi _4^2+2 \varphi _2^2 (\varphi _2^2\nonumber\\&-&2 \varphi _5^2)
   \varphi _7^2] \varphi _1^5+32 \varphi _2 \varphi _4 \varphi _5 [\left(11 \varphi _7^2-5 \varphi _4^2\right) \varphi _2^2+\varphi _5^2 \left(2 \varphi
   _4^2-7 \varphi _7^2\right)] \varphi _1^4\nonumber\\&+&32 \varphi _4^2 [\left(\varphi _4^2+\varphi _7^2\right) \varphi _2^4+5 \varphi _5^2 \left(\varphi _4^2+2
   \varphi _7^2\right) \varphi _2^2-3 \varphi _5^4 \left(\varphi _4^2+\varphi _7^2\right)] \varphi _1^3\nonumber\\&+&32 \varphi _2 \varphi _4^3 \varphi _5 [\varphi
   _4^2 \left(5 \varphi _5^2-4 \varphi _2^2\right)-2 \left(\varphi _5^2-5 \varphi _2^2\right) \varphi _7^2] \varphi _1^2-32 \varphi _4^4 [\varphi _7^2
   \varphi _2^4\nonumber\\&+&\varphi _5^2 \left(\varphi _4^2-14 \varphi _7^2\right) \varphi _2^2+3 \varphi _5^4 \varphi _7^2] \varphi _1-32 \varphi _2 \varphi _4^5 \varphi
   _5 [\varphi _2^2 \left(\varphi _4^2+\varphi _7^2\right)-5 \varphi _5^2 \varphi _7^2]]\nonumber\\&/&b(t) \, dt]/[96 \left(\varphi _2 \varphi _4-\varphi _1 \varphi
   _5\right){}^2 \left(\varphi _1 \varphi _2+\varphi _4 \varphi _5\right){}^2]+\chi _6,\nonumber\\
   \varphi_6(t)&=&[\left(\varphi _1^2+\varphi _4^2\right) [\int [\left(\varphi _2 \varphi _4-\varphi _1 \varphi _5\right){}^2 [\varphi _1 \varphi _5 \varphi _2
   \left(c(t)+\varphi _1^2+3 \varphi _4^2\right)\nonumber\\&+&\varphi _4 \varphi _5^2 \left(c(t)-\varphi _4^2\right)+3 \varphi _1^2 \varphi _4 \varphi _2^2]+\left(\varphi
   _1 \varphi _2+\varphi _4 \varphi _5\right) [\left(5 \varphi _2^2 \varphi _5-\varphi _5^3\right) \varphi _1^2\nonumber\\&-&3 \varphi _2 \varphi _4 \left(\varphi _2^2-3
   \varphi _5^2\right) \varphi _1+2 \varphi _4^2 \varphi _5 \left(\varphi _5^2-2 \varphi _2^2\right)] \varphi _7^2]/b(t) \, dt]]\nonumber\\&/&[3 \left(\varphi _2
   \varphi _4-\varphi _1 \varphi _5\right){}^2 \left(\varphi _1 \varphi _2+\varphi _4 \varphi _5\right){}^2]+\chi
   _7,\nonumber\\
   \varphi_9(t)&=&[\int [\left(\varphi _2 \varphi _4-\varphi _1 \varphi _5\right){}^2 \left(\varphi _1 \varphi _2+\varphi _4 \varphi _5\right) [\varphi _4 \varphi _5
   \varphi _1^2+\varphi _2 \left(\varphi _7^2-\varphi _4^2\right) \varphi _1\nonumber\\&+&\varphi _4 \varphi _5 \varphi _7^2] c(t)+\left(\varphi _1 \varphi _2+\varphi _4
   \varphi _5\right){}^2 [\left(2 \varphi _2^2-\varphi _5^2\right) \varphi _1^2+6 \varphi _2 \varphi _4 \varphi _5 \varphi _1\nonumber\\&-&\varphi _4^2 \left(\varphi _2^2-2
   \varphi _5^2\right)] \varphi _7^4-\left(\varphi _2 \varphi _4-\varphi _1 \varphi _5\right) \left(\varphi _1 \varphi _2+\varphi _4 \varphi _5\right)
   [\varphi _4 (\varphi _1^3\nonumber\\&-&\varphi _1 \varphi _4^2) \varphi _2^2+4 \left(\varphi _1^4+3 \varphi _4^2 \varphi _1^2+\varphi _4^4\right) \varphi _5
   \varphi _2+\varphi _1 \varphi _4 \left(\varphi _4^2-\varphi _1^2\right) \varphi _5^2] \varphi _7^2\nonumber\\&+&\varphi _1 \varphi _4 \left(\varphi _1 \varphi
   _5-\varphi _2 \varphi _4\right){}^3 \left(\varphi _1^3 \varphi _2-\varphi _4^3 \varphi _5\right)]/b(t) \, dt]\nonumber\\&/&[3 \left(\varphi _2 \varphi _4-\varphi _1 \varphi
   _5\right){}^2 \left(\varphi _1 \varphi _2+\varphi _4 \varphi _5\right){}^2 \varphi _7]+\chi
   _8, \theta _3(t)=\chi
   _9 \sqrt{\theta _1(t)},\nonumber\\ \theta _2(t)&=&\chi
   _{10} \sqrt{\theta _1(t)}, a(t)=-\frac{3 \left(\varphi _1 \varphi _2+\varphi _4 \varphi _5\right) b(t)}{\varphi _1^2+\varphi
   _4^2},\nonumber\\ \chi_{10}&=&\epsilon_2 \sqrt{\varphi _1^2+\varphi _7^2} \sqrt{\frac{\chi _9^2}{\varphi _1^2+\varphi _4^2}-\frac{4}{\varphi _4^2-\varphi _7^2}}
\end{eqnarray}where  $\chi _i (i=6,7,8,9,10)$ is integral constant,  $\epsilon_2=\pm1$.\\

\noindent {\bf Appendix C}\\ \begin{eqnarray}(I)\,\,\, k(t)&=&[a(t)
b(t) c(t)-[3 \left(\alpha _1^2+\alpha _4^2\right) \left(\alpha _1
\alpha _2+\alpha
   _4 \alpha _5\right) [2 \left(\alpha _1 \alpha _2+\alpha _4 \alpha _5\right) a(t)
   b(t)\nonumber\\&+&\left(\alpha _1^2+\alpha _4^2\right) a(t)^2+\left(\alpha _2^2+\alpha _5^2\right)
   b(t)^2]]/[\left(\alpha _2 \alpha _4-\alpha _1 \alpha _5\right){}^2 \alpha _7(t)]]/a(t)^2,\nonumber\\
   \alpha_3(t)&=&\int [[3 \left(\alpha
_1^2+\alpha _4^2\right) \left(\alpha _1 \alpha _2+\alpha _4 \alpha
   _5\right) [[2 \alpha _2 \alpha _4 \alpha _5+\alpha _1 \left(\alpha _2^2-\alpha
   _5^2\right)] a(t)\nonumber\\&+&\alpha _2 \left(\alpha _2^2+\alpha _5^2\right) b(t)]]/[\left(\alpha
   _2 \alpha _4-\alpha _1 \alpha _5\right){}^2 \alpha _7(t)]-\alpha _2 a(t) c(t)]/a(t)^2 \,
   dt,\nonumber\\
\alpha_6(t)&=&\int [[3 \left(\alpha _1^2+\alpha _4^2\right)
\left(\alpha _1 \alpha _2+\alpha _4 \alpha
   _5\right) [\left(-\alpha _4 \alpha _2^2+2 \alpha _1 \alpha _5 \alpha _2+\alpha _4 \alpha
   _5^2\right) a(t)\nonumber\\&+&\alpha _5 \left(\alpha _2^2+\alpha _5^2\right) b(t)]]/[\left(\alpha _2
   \alpha _4-\alpha _1 \alpha _5\right){}^2 \alpha _7(t)]-\alpha _5 a(t) c(t)]/a(t)^2 \, dt,\nonumber\\ \alpha_7(t)&=&\chi_{11},
\end{eqnarray}
\begin{eqnarray}(II)\,\,\, \alpha_5&=&\alpha_2=0, \alpha_7(t)=\int -\frac{2 \left(\alpha _1^2+\alpha _4^2\right) \alpha _3(t) \int \alpha _3(t) \,
   dt}{\alpha _4^2} dt,\nonumber\\
\alpha_6(t)&=&-\frac{\alpha _1 \alpha _3(t)}{\alpha _4},
\end{eqnarray}
\begin{eqnarray}(III)\,\,\, k(t)&=&\frac{b(t) c(t)}{a(t)},  \alpha_5=-\frac{\alpha _1 \alpha _2}{\alpha _4}, \alpha_7(t)=\chi_{12},\nonumber\\
   \alpha_3(t)&=&-\frac{\alpha _2 c(t)}{a(t)},
\alpha_6(t)=\frac{\alpha _1 \alpha _2 c(t)}{\alpha _4 a(t)},
\end{eqnarray}
\begin{eqnarray}(IV)\,\,\, a(t)&=&0,  \alpha_5=-\frac{\alpha _1 \alpha _2}{\alpha _4}, \alpha_7(t)=\chi_{13},
\nonumber\\
   \alpha_3(t)&=&-\frac{\alpha _1 c(t)+\alpha _2 k(t)}{b(t)}, \alpha_6(t)=\frac{\alpha _1 \alpha _2 k(t)-\alpha _4^2 c(t)}{\alpha _4 b(t)},
\end{eqnarray}where  $\chi _i (i=11,12,13)$.

\end{document}